# Automating the GDPR Compliance Assessment for Cross-border Personal Data Transfers in Android Applications**

Danny S. Guamán, Xavier Ferrer, Jose M. del Alamo, Jose Such

**Abstract**— The General Data Protection Regulation (GDPR) aims to ensure that all personal data processing activities are fair and transparent for the European Union (EU) citizens, regardless of whether these are carried out within the EU or anywhere else. To this end, it sets strict requirements to transfer personal data outside the EU. However, checking these requirements is a daunting task for supervisory authorities, particularly in the mobile app domain due to the huge number of apps available and their dynamic nature. In this paper, we propose a fully automated method to assess compliance of mobile apps with the GDPR requirements for cross-border personal data transfers. We have applied the method to the top-free 10,080 apps from the Google Play Store. The results reveal that there is still a very significant gap between what app providers and third-party recipients do in practice and what is intended by the GDPR. A substantial 56% of analysed apps are potentially non-compliant with the GDPR cross-border transfer requirements.

──────────────── ◆ ────────────────

## 1 INTRODUCTION

THE distributed nature of today's digital systems and services not only facilitates the collection of personal data from individuals anywhere, but also their transfer to different countries around the world [1]. This raises potential risks to the privacy of individuals, as the organizations sending and receiving personal data can be subject to different data protection laws and, therefore, may not offer an equivalent level of protection. In some regions, such as China, privacy may be less valued, or valued differently, when compared to order and governance [2]. While in other regions, particularly in the European Union (EU), privacy is strenuously protected [3] and is conceived as a Human Right [4]. As a result, the General Data Protection Regulation (GDPR) [3], constraints cross-border transfers (*also named international transfers*) outside the European Economic Area (EEA)[1] and recognises only twelve non-EU countries as providing protection equivalent to the GDPR.

Mobile applications, or just "apps", exacerbate the data protection compliance issues for organizations, notably with requirements related to cross-border transfers. The particularities of the app development and distribution ecosystems are major factors underlying these issues [5]. First, apps collect a great amount of personal data, which may be transmitted from the device to data processors across the world [1], or shared between chains of third-party service providers [6], even without the app developer's knowledge [7]. Second, apps are distributed through global stores, enabling app providers to easily reach markets and users beyond its country of residence. In this context, there is a need for constant vigilance by the various stakeholders, including app developers, supervisory authorities, and app distribution platforms, to ensure that appropriate requirements have been met and to avoid potential data protection compliance breaches.

Nevertheless, testing or auditing mobile apps against legal data protection requirements is challenging. First, it is necessary to simplify the high-level legal requirements into concrete step-by-step technical criteria and indicators to be assessed in mobile apps. Second, parties responsible for checking compliance with data protection requirements, such as supervisory authorities, require automated assessment methods and tools to cope with the vast and ever changing mobile ecosystem [5]. Even distribution platforms like Google Play Store can benefit from these automated assessment methods and thus extend their protection mechanisms, such as Google Play Protect, which currently focuses on detecting potential harmful behaviour in the Android ecosystem only from the security perspective.

Our main contributions to address this challenge are:

1) An automated method to assess compliance of Android apps with the cross-border transfer requirements established by the GDPR. It leverages our prior work on a compliance assessment process [8], and further extends it with an automated approach to identify cross-border transfer statements from natural language privacy policies. With an F-measure ranging from 85.7% to 100% in identifying the different cross-border transfer statements,

────────────

** Author's copy of the manuscript
- *Danny S. Guamán is with Universidad Politécnica de Madrid, Spain and Escuela Politécnica Nacional, Ecuador. E-mail: danny.guaman@epn.edu.ec.*
- *Xavier Ferrer is with the Department of Informatics, King's College London, United Kingdom.*
- *Jose M. del Alamo is with Universidad Politécnica de Madrid, 28040 Madrid, Spain. E-mail: jm.delalamo@upm.es*
- *Jose Such is with the Department of Informatics, King's College London, United Kingdom. E-mail:*

[1] The EEA includes all the EU Member States plus Norway, Iceland and Liechtenstein. For the sake of clarity, we will use the term EU from now on to refer to all of these countries.

our approach can be exploited to extract these privacy practices with a high degree of certainty.

2) A large-scale assessment of Google Play Store apps compliance with the cross-border transfer requirements of GDPR. We leveraged our automated method to assess the top-free 10,080 apps in Spain and the top 110 third-party services they use.

## 2 RELATED WORK

The automated compliance assessment of Android apps requires the analysis of both the privacy policy text and the app behaviour.

For the privacy policy analysis, automated approaches [9] [10] rely on the codification or annotation method [11], where one or multiple domain analysts generate structured annotations of privacy practices (*i.e.,* a corpus) by systematically assigning a label to the policy statements. Useful corpora have been released in the privacy domain [10], [12]. These corpora ultimately are used as ground truth for building automatic classification models. For example, Zimmeck et al. [10] automated the extraction of data collection practices from privacy policies, while Andow et al. [13] distinguished the entity (*i.e.* first-party vs. third-party) to which personal data is sent.

Focusing on GDPR, Fan et al. [14] empirically assessed transparency, data minimization, and confidentiality requirements in Android mHealth apps, checking whether six different practices are informed through privacy policies. Mangset [15] also checked GDPR requirements related to transparency, data minimization (collection practices), confidentiality (data at rest in transit), and some user rights (particularly, consent and objection automatically individual decision-making). Unfortunately, none of these works addressed cross-border transfer practices.

As for the app behaviour analysis, researchers have leveraged static, dynamic or hybrid techniques. Ferrara and Spoto [16] relied on static code analysis to detect disclosures of personal data so that data protection officers could spot potential GDPR infringements. Jia et al. [17] leveraged dynamic techniques to detect personal data disclosures in network packets lacking user consent. While our work focuses on different GDPR requirements, Jia's work could be seen as complementary to our work as the app behavoiur analysis method could minimize the false-negative rate.

Finally, we consider Eskandari et al. [18] as the closest related work with regard to the GDPR requirements covered. They propose PDTLoc, an analysis tool that employs static analysis to detect violations of article 25.1 of the EU Data Protection Directive (European data protection law replaced by the GDPR). This Directive set requirements for international transfers similar to those laid down in the GDPR. However, this prior work presumes any transfer outside the EU to be a regulatory infringement, thus this approach would have incorrectly identified potential compliance issues. The authors did not consider the privacy policies as a means of disclosing the intention to perform cross-border transfers and the appropriate safeguards that do enable these transfers.

To fill this gap, in prior work [8] we defined an earlier method for the compliance assessment of Android apps with GDPR cross-border personal data transfer requirements. This work supported the app behaviour analysis through dynamic testing techniques. It also identified the specific requirements for the transparency elements to be included in the privacy policies for the lawful disclosure of the international transfer. However, the compliance assessment process was not automated, as the interpretation of the privacy policies required human analysis, and thus did not scale. We have extended this prior work with an automated approach to identify cross-border transfer statements from natural language policies. As a result, we have been able to carry out an extensive assessment of cross-border transfers in Google Play Store apps.

## 3 GDPR CROSS-BORDER TRANSFERS

As illustrated in Figure 1, there are specific criteria for determining what must be considered a cross-border transfer, to which the GDPR lays down further requirements. These requirements include, in each case, the disclosure of specific and meaningful information to data subjects. Next, we briefly summarize the criteria, and refer the interested reader to [8] for details.

The **criterion C1.1** determines whether personal data, in the meaning of GDPR [19], are sent to remote recipients (See DT enumeration in Fig. 1).

The **criterion C1.2** determines whether an app targets EU citizen. We fairly assume that mobile apps available in the Google Play Store reachable from an EU country are indeed targeting EU users.

The **criterion C1.3** determines to which country personal data are sent. Data transfers between EU countries do not add further constraints or requirements.

The **criterion C1.4** distinguishes whether the servers located outside the EU belong to the app provider itself (*i.e. first-party recipient* or data controller in GDPR terms), or another organization (*i.e. third-party recipient*). In the former case, if the first-party recipient is also located outside the EU then it must disclose the contact details of its representative in the EU[2] (T1 in Fig. 1). The latter is considered an international data transfer.

The **criterion C1.5** seeks to determine the specific transparency requirements of the international transfer, namely adequacy or non-adequacy decision.

Twelve non-EU countries have been recognized as providing data protection equivalent to the GDPR and therefore maintaining an *adequacy decision* (See ADC enumeration in Fig. 1). International transfers to these countries can take place without any further safeguards. However, to ensure transparency[3] the app provider should disclose (1) the *intention* to transfer personal data to a non-EU country, (2) the names of targeted countries, and (2) the *existence* of an adequacy decision by the Commission [20] (T2 in Fig. 1).

---

[2] GDPR Art. 27(1)
[3] GDPR Art. 13(1)(f) and Art. 14(1)(f)

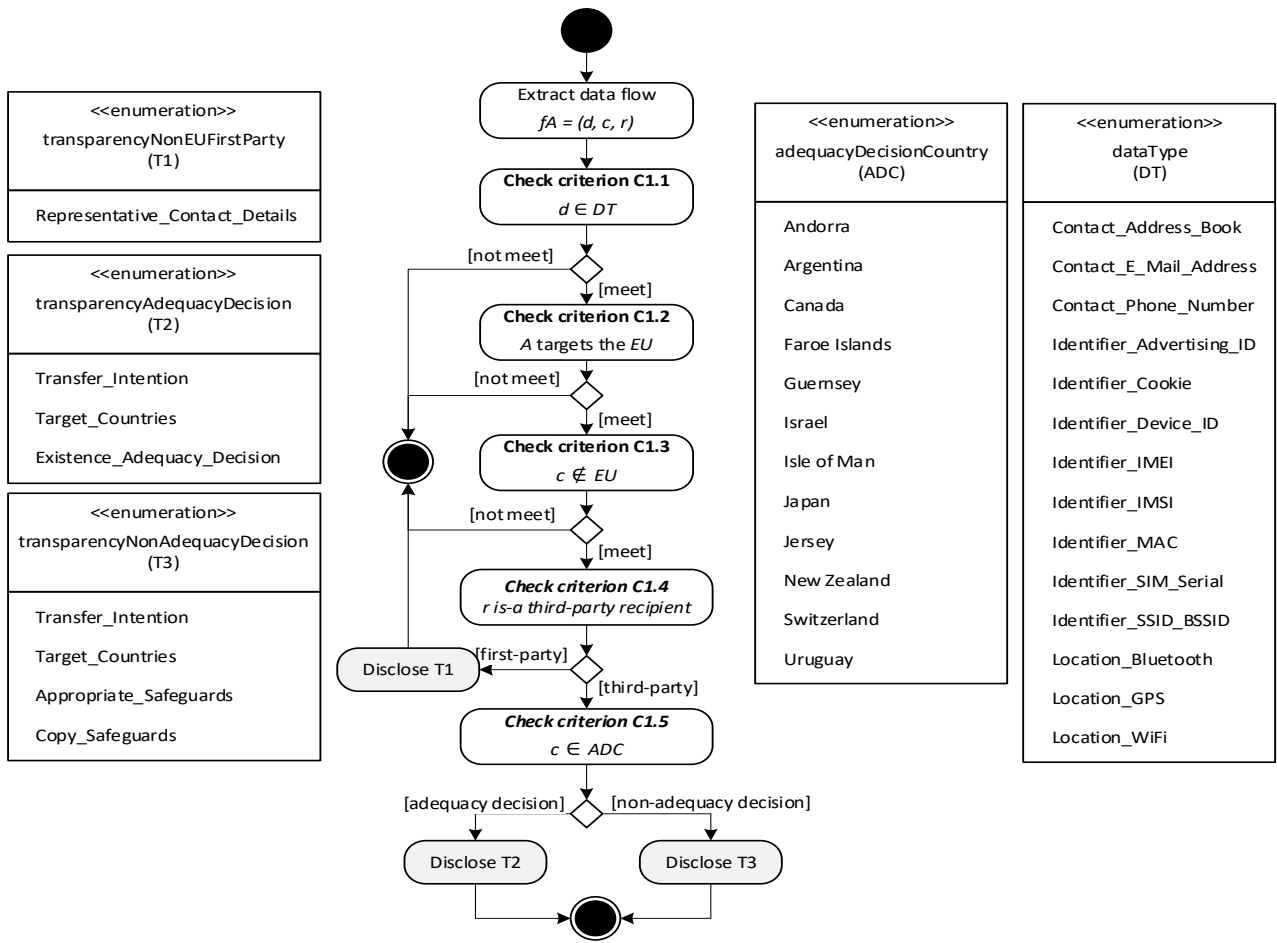

Fig. 1. Criteria for distinguishing the type of cross-border transfer performed by each app (A) targeting European Union (EU) data subjects. Each flow of mobile application A (fA) is represented by the type of personal data (d), destination country (c), and type of recipient (r). In each case (grey boxes), specific information (Ti) should be disclosed to data subjects, generally through privacy policies to ensure transparency.

International transfers not covered by an adequacy decision requires app providers to adopt *"appropriate safeguards"* before carrying them out. GDPR defines assurance mechanisms to enable international transfers in such cases[4], including *Standard Data Protection Clauses, Binding Corporate Rules, Approved Codes of Conduct* and *Approved Certification Schemes*. These assurance mechanisms should be approved by the EU and, in general, allow the app provider to ensure that third-party recipients have implemented appropriate safeguards to guarantee a protection level equivalent to GDPR. The EU has adopted four Standard Data Protection Clauses [21], which should be incorporated into contracts between the parties concerned. Binding Corporate Rules should be established when international transfers take place between companies belonging to a corporate group and should be signed with the Commission's approval. To the best of our knowledge, the EU has not yet adopted any Code of Conduct or Certification Scheme for GDPR. Finally, to ensure transparency, the app provider should inform the data subjects about (1) the *intention* to transfer personal data to a non-EU country, (2) the names of targeted countries, (3) a reference to the *appropriate safeguard(s)* according to the aforementioned options, and (4) the means to obtain a copy of the safeguard(s) [20] (T3 in Fig. 1).

In the absence of an *adequacy decision* or any *appropriate safeguards*, some exceptions[5] allow for international transfers in specific situations. We highlight the *explicit consent*, which requires the consent through an affirmative action of the data subjects, *e.g.*, ticking a box, to be obtained after providing precise details of the international transfers. In this case, the data subject should also be able to withdraw consent easily at any time.

## 4 COMPLIANCE ASSESSMENT METHOD

Assessing compliance of an app with cross-border transfers requires fundamentally three activities (*privacy policy analysis, app behaviour analysis* and *compliance checking*) and two inputs (the privacy policy used for the *privacy policy analysis* and the Android application package APK used for the *App analysis*) as shown in Fig. 2.

The *privacy policy analysis* parses the privacy policy of the app to extract the cross-border transfer practices disclosed by the app provider.

---

[4] GDPR Chapter V  
[5] GDPR Art. 49

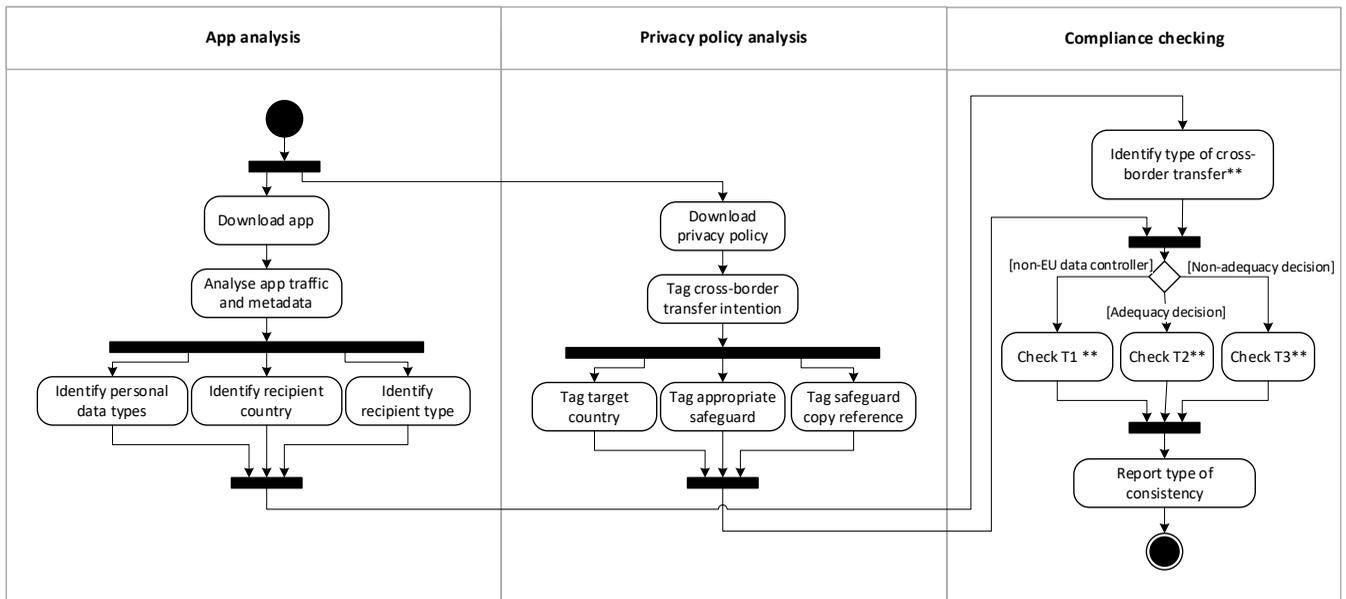

Fig. 2. The automated pipeline that detects the app's cross-border transfers and automatically checks them against the corresponding privacy practices in its privacy policy. (**) These activities have been detailed in Fig. 1

In parallel, *the app analysis* part consists of installing and executing the application (APK) to observe its real behaviour, particularly the type of personal data it leaks, the type of recipient who receives the data (i.e., first-party or third-party recipient), the country in which the recipient servers are located, and the information on the app digital certificate that determines the app provider and its location.

Finally, based on the practices disclosed through the app privacy policy and the cross-border transfers it actually performs, a *compliance checking* alerts of potential non-compliant behaviour.

Section 4.1 details the *privacy policy analysis*, while Sections 4.2 and 4.3 summarise the relevant aspects of the *app analysis* and *compliance checking*, respectively. For more details on the latter two, refer to [8].

## 4.1 Automated Privacy Policy Analysis

In this section, we present our automated approach for classifying cross-border transfer practices in privacy policies (Table 1). We relied on the IT-100 Corpus[6] that consists of one hundred privacy policies manually annotated by two privacy experts and contains 3,715 policy segments of which 281 segments contain transparency elements of cross-border transfers.

As illustrated in Fig. 3, individual transparency elements are part of an entire policy segment disclosing a cross-border transfer practice. Therefore, we composed a two-layer classification pipeline: a cross-border transfer intention classifier to identify entire policy segments disclosing the intention to perform a cross-border transfer (Section 4.1.1), followed by a transparency element classifier to identify the individual transparency elements disclosed (Section 4.1.2). The validation of the two-layer classification pipeline is presented in Section 4.1.3.

---

[6] The corpus has been released at https://github.com/PrivApp/IT100-Corpus

TABLE 1. CROSS-BORDER TRANSFER ANNOTATION SCHEME.

| Cross-border transfer type | Required transparency elements |
|---|---|
| T1. Transfer to non-EU data controller | EU Representative contact information |
| T2. International transfer (with adequacy decision) | Transfer intention |
| | Existence of EU adequacy decision |
| | Target country |
| T3. International transfer (without adequacy decision) | Transfer intention |
| | Target country |
| | Appropriate safeguards:<br>-Standard Data Protection Clauses<br>-Binding Corporate Rules<br>-Approved Codes of Conduct<br>-Approved Certification Schemes<br>-Explicit consent |
| | Copy means |

### 4.1.1 Cross-border transfer intention classifier

This classifier tags each policy segment, roughly a paragraph, to indicate whether it discloses (1) or not (0) the intention to perform a cross-border transfer. We use the IT-100 corpus to train a supervised machine learning (ML) algorithm and generate a binary classifier. We have followed the systematic process shown in Figure 4 to determine the best performance classification model.

> Our business may require us to transfer your Personal Data to countries outside of the European Economic Area (EEA), including to countries such as the Peoples Republic of China or Singapore. We take appropriate steps to ensure that recipients of your Personal Data are bound to duties of confidentiality and we implement measures such as standard contractual clauses. A copy of those clauses can be obtained by contacting our Help

Fig. 3. Segments of the privacy policy of the net.manga.geek.mangamaster app. The segment discloses a typical cross-border transfer practice, including the transfer intention (blue), the target country (red), the appropriate safeguards (purple), and the means to get a copy of such safeguards (green).

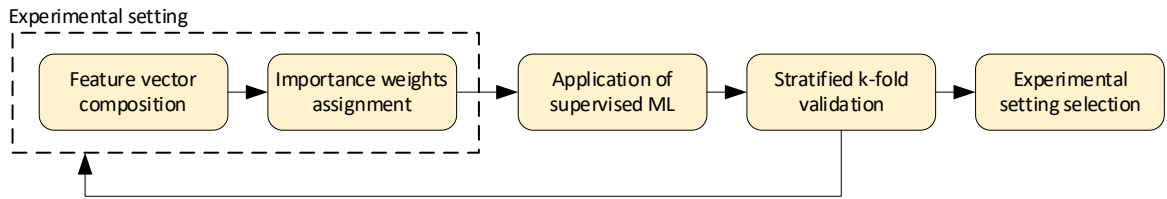

Fig. 4. The overall process to generate the cross-border transfer classification model.

a) **Feature vector composition.** We relied on the *bag-of-words* model to define a set of candidate features based on all distinct terms in the IT-100 corpus policy segments. We tokenized the 3,715 policy segments and extracted a set of 9,239 candidate features. The rationale for this key assumption is that there is a distribution of individual terms in cross-border transfer practices that is distinct from the distribution of unrelated practices. Figure 5 depicts the relative document frequency of the 30-top terms[7] that mostly appear in segments (1) disclosing cross-border transfers practices and (2) disclosing other unrelated practices (*e.g.*, collection or sharing practices). For illustrative purposes, we represent the 30-top terms of each segment class in the same plot, which have then been rearranged to distinguish the terms frequently used in both segment classes and the terms frequently used in one of them. There is a subset of generic terms which are frequently used in both segment classes. For example, the term *"information"* appears in 64% of the cross-border transfer segments and also in 58% of the unrelated segments. The same applies to other terms such as *"privacy"*, *"data"* or *"protection"*, which are generic and are not tied to a specific privacy practice. On the other hand, a subset of terms is related to cross-border transfers (*e.g.*, *"transfer"*, *"country"*, and *"outside"*). These terms appear mostly in the segments disclosing cross-border transfers and only marginally in unrelated segments. For example, the term *"transfer"* appears in 87% of the cross-border transfer statements but only in 4% of the other statements.

We pre-processed the 9,239 candidate features, removing number-related tokens, punctuation, non-ASCII characters, *stop words*, and duplicated tokens after normalizing them to lowercase. This pre-processing removed up to 32% of features worthless in distinguishing cross-border transfers, leaving a total of 6,204 features.

Furthermore, while individual terms of policy segments (*e.g.*, "*transfer*") are relevant for categorizing cross-border transfer practices, other relevant term units built upon a contiguous sequence of *n* terms can also be relevant (*e.g.*, "*approved contractual clauses*"). Therefore, apart from the individual terms, we relied on the *n*-gram model to parse each policy segment into new composite features. We experimented empirically to determine the value of *n*, selecting the range of *n*-grams that provides the highest performance metrics.

b) **Importance weight assignment.** We experimented with three different approaches to assigning weights to the selected features: a binary counter (BC), the term frequency (TF) and the Term Frequency-Inverse Document Frequency ($TF.IDF$). The BC encodes the presence or absence of each feature. The $TF$ encodes the number of times that each feature occurs in a policy segment. Third, the $TF.IDF$ relies on the $TF$ to encode the number of times that each feature occurs in a policy segment, but the $IDF$ penalizes (decreases) it as the feature $x_i$ occurs across many policy segments. Formally, the $TF.IDF$ for the feature $x_i$ is computed as $TF_i . log \frac{N}{n_i}$, where $N$ is the total number of IT-100 corpus policy segments, and $n$ is the number of policy segments that contain the feature $x_i$.

c) **Application of an ML supervised algorithm.** The above-mentioned feature vectors and their corresponding class labels can be denoted as $S = \{(x_1, y_1), ..., (x_n, y_n)\}$, where $x_i$ is the feature vector of the policy segment $i$, and $y_i \in \{0,1\}$ indicates the class label of the policy segment. In our case, 1 implies the policy segment $i$ discloses the intention to perform a cross-border transfer, and 0 its absence. By using the training sample $S$, stemmed from the IT-100 corpus, we used the Support Vector Machine (SVM) technique to find the optimal separation hyperplane that best divides the dataset into the two classes mentioned. SVM has been empirically demonstrated better performance over a variety of other ML techniques in high dimensional spaces, being still effective in cases where the number of dimensions is greater than the number of samples[22]. Also, prior work [9], [23] demonstrated that SVM can reach higher performance than Logistic Regression and Convolutional Neural Network for privacy practices classification.

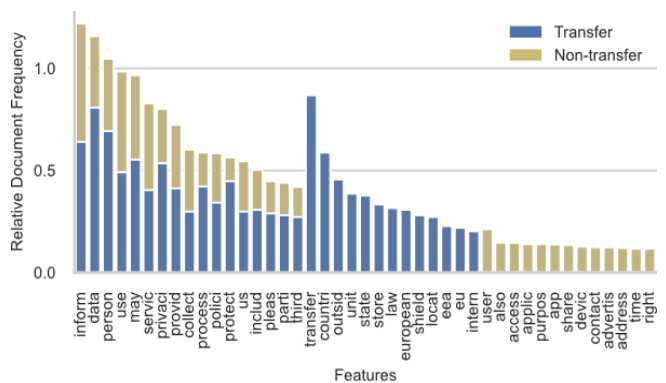

Fig. 5. Relative document frequency of top-30 terms that appear in cross-border transfer segments (blue) and other unrelated segments (yellow).

---

[7] Before computing the relative document frequency of each term, we removed morphological affixes from terms and took their roots by using the Porter Snowball Stemmer [25] in order to resolve the usage of inflected words. The full relative frequency distribution can be found in the replication package available at http://dx.doi.org/10.17632/drx5nc3hr4

**d) Stratified k-fold validation.** The IT-100 corpus is an imbalanced dataset *i.e.* the number of samples tagged as disclosing cross-border transfers is much lower than the number of samples not disclosing it. Accordingly, stratified *k*-fold cross-validation was carried out, establishing $k = 5$. The advantage of the stratified *k*-fold cross-validation is that the entire dataset is used for both training and testing while ensuring that policy segments tagged as disclosing a cross-border transfer practice are represented consistently among all training and validation folds.

**e) Results and experimental setting selection.** We performed an empirical evaluation of the effect of iteratively applying the different experimental settings explained above (Table 2). We relied on Scikit Learn [24] and the Natural Language Toolkit [25] to carry out all our experiments. In each case, we computed the standard performance metrics to compare the models generated. Since the IT-100 corpus is an imbalanced dataset, we primarily used the F-measure instead of the accuracy metric. For models with comparable F-measures, we have favoured the model with the highest recall for the sake of more conservative analysis. That is, we seek to prevent a privacy policy that does disclose the intention to perform a cross-border transfer from being tagged as not doing so.

TABLE 2. SETTING PARAMETERS USED TO BUILD THE CROSS-BORDER CLASSIFICATION MODELS.

| Task | Approaches |
|---|---|
| Feature vector generation | Individual terms |
| | N-gram terms (1-4) |
| | Stemmed N-gram terms |
| Importance weight assignment | Binary counter |
| | Term frequency (TF) |
| | Term frequency-Inverse Document Frequency (TF.IDF) |

Table 3 summarises the results. The experimental setting that consistently provided better performance (F-measure of 90.9%) built on top of a feature vector of stemmed uni- and bigrams and TF as the weighting assignment approach. Also, the best SVM parameters are the *Modifier-Huber loss function* and *SVM alpha* of $10^{-3}$. This setting has been used to build the definitive binary classification model to identify the entire policy segments disclosing the intention to perform a cross-border transfer. This is then placed at the entry to the transparency element classifiers explained in the next section.

### 4.1.2 Transparency elements classifiers

The policy segments disclosing the intention to perform cross-border transfers are further processed to identify specific transparency elements, as per Table 1.

**Target country classification.** After analysing the IT-100 Corpus, we observed that target countries are disclosed in three different ways. First, explicit country names or their abbreviations (*e.g.* U.S.) are mostly disclosed in privacy policies. Second, some domain-specific terms were also used to implicitly disclose the target countries. For example, *Privacy Shield* is a certification framework that, until 16 July 2020, ensured an adequacy decision to perform international transfers to the United States. Third, city names rather than country names are also used by a minor number of privacy policies. Our approach, therefore, involves a dictionary of countries, cities and aliases, whose occurrences are sought in the pertinent policy segments. More specifically, we relied on the CountryInfo dataset[8] that provides details on all countries, including their canonical names, country codes, as well as their states and provinces. We extended it by adding an '*alias*' field to register domain-specific terms implicating a particular country. For example, "*Privacy Shield*" was added as an alias for the United States. Both policy segments disclosing a cross-border transfer and the country dictionary values were first normalised to lowercase. Then, if a non-EU country, state, province, or alias from the dictionary occur in a policy segment, it is labelled with the identified country/s.

The results of applying this approach to a set of 117 IT-100-Corpus policy segments disclosing cross-border transfers are shown in Table 4. Overall, we observe that the approach provides high performance in detecting target countries in policy segments. Admittedly, the number of positive ground truth is minimum for certain countries, but even in the case of the United States, which is the most mentioned country, the method works strongly despite being disclosed in different ways. We observed that a couple of policy segments were misclassified due to typos in the country names or compound ways of referring to countries (*e.g. California-based*), which scape the proposed approach. On the other hand, this approach depends directly on the policy segments fed by the binary classifier of cross-border transfers.

TABLE 3. PERFORMANCE OF CROSS-BORDER CLASSIFICATION MODELS BY USING DIFFERENT WEIGHT ASSIGMENTS.

| Weighting approach | N-gram | Precision | Recall | F-measure |
|---|---|---|---|---|
| BC | 1 | 81.6% | 80.8% | 81.2% |
| | 1-2 | 83.4% | 75.2% | 79.1% |
| | 1-3 | 87.0% | 64.8% | 74.3% |
| | 1-4 | 87.7% | 57.6% | 69.5% |
| TF | 1 | 90.1% | 90.4% | 90.2% |
| | 1-2 | 89.9% | 92.0% | 90.9% |
| | 1-3 | 83.0% | 94.4% | 88.3% |
| | 1-4 | 76.6% | 94.4% | 84.5% |
| TF-IDF | 1 | 87.7% | 88.0% | 87.9% |
| | 1-2 | 87.9% | 88.8% | 88.3% |
| | 1-3 | 85.0% | 89.6% | 87.2% |
| | 1-4 | 84.7% | 90.4% | 87.5% |

All these classification models were trained by considering stemmed features and different n-gram sizes. The experimental settings that got the highest performance metric are distinguished in blue, while the lowest performance metrics are distinguished in red.

**Appropriate *safeguard and copy means*.** We built one binary SVM classifier (*Adequacy Decision*) and four keyword-based rule classifiers (*Standard Data Protec-*

---
[8] https://pypi.org/project/countryinfo

tion

TABLE 4. PERFORMANCE OF TARGET COUNTRY CLASSIFIER.

| Country | Precision | Recall | F-measure | +/- Support |
|---|---|---|---|---|
| Japan | 100% | 100% | 100% | 2/115 |
| South Korea | 100% | 100% | 100% | 1/116 |
| United Arab Emirates | 100% | 100% | 100% | 1/116 |
| Australia | 100% | 100% | 100% | 3/114 |
| Singapore | 100% | 100% | 100% | 6/111 |
| India | 100% | 100% | 100% | 1/116 |
| United States | 100% | 97% | 99% | 71/46 |
| China | 100% | 100% | 100% | 5/112 |
| Canada | 100% | 100% | 100% | 1/116 |

It builds upon a set of policy segments ($n = 117$) that disclose the intention to perform a cross-border transfer. The Support column shows the number of ground truths of (+) the segments that disclose the corresponding target country and (-) the segments that do not disclose any target country.

Clauses, Binding Corporate Rules, Explicit Consent, and Means to get a copy of safeguards) to identify the other individual transparency elements besides the target country. We have not generated classifiers for Approved Certification and Approved Code of Conduct as they are not disclosed in any IT-100 Corpus privacy policy. This makes sense since, so far, the Commission has not yet adopted any Code of Conduct or Certification Scheme for GDPR.

The binary SVM classifier to identify cross-border transfer statements covered by an Adequacy Decision was built by following the same procedure explained in Section 3.1.1. The best performance was achieved by a binary SVM classifier built on top a uni- and bigram-based features, and using a TF.IDF weighting approach. The details on the performance achieved by other classifiers we tried are available in the accompanying replication package.

On the other hand, due to the limited number of positive ground truths and the higher performance compared to binary classifiers, we generated keyword-based rule classifiers to identify the remaining four transparency elements. Our approach involves developing a set of rules leveraging that a scoped domain-specific vocabulary is used to refer them. The set of 117 IT-100-Corpus policy segments disclosing a cross-border transfer were analysed by a privacy expert, who selected minimum phrases (2-5 terms) that captured the key terms of each transparency element. These phrases were first normalized to lowercase and their grammatical roots by using Porter Stemmer and then turned into a set of rules. For example, the rule ('contract'|'standard') w/4 ('model'|'clause') implies that the normalized form 'contract' or 'standard' must occur before or after the normalised form 'model' or 'clause' by no more than 4 terms in the same sentence. If that rule is satisfied, a policy segment is labelled as disclosing Standard Data Protection Clauses. The same procedure was followed for the other transparency elements.

The results of applying these classifiers to a set of 117 IT-100-Corpus policy segments disclosing cross-border transfers are shown in Table 5. The binary adequacy decision classifier achieves high performance, only misclassifying policy segments ambiguously stated. The keyword-based rule classifiers also allowed to correctly identify most of the disclosed transparency elements in the IT-100 Corpus policy segments. Admittedly, their generalization may be hindered since their rules were built based on a relatively small positive ground truth. We, therefore, performed a further evaluation on a subset of unseen privacy policies as explained next.

TABLE 5. PERFORMANCE OF IDENTIFYING THE TRANSPARENCY ELEMENTS ON A SET OF POLICY SEGMENTS ($n = 117$) THAT DISCLOSE THE INTENTION TO PERFORM A CROSS-BORDER TRANSFER.

| Transparency element | Precision | Recall | F-measure | +/- Support |
|---|---|---|---|---|
| Adequacy Decision | 97% | 90% | 94% | 41/76 |
| Standard Data Protection Clauses | 100% | 100% | 100% | 12/105 |
| Binding Corporate Rules | 100% | 100% | 100% | 4/113 |
| Explicit Consent | 60% | 100% | 75% | 4/113 |
| Copy Reference | 100% | 100% | 100% | 5/112 |

The Support column shows the number of ground truths of (+) the segments that disclose the corresponding transparency element and (-) the segments that do not disclose it. Adequacy decision results relied on 3-fold stratified cross-validation.

4.1.3 Validation

We validated the two-layer classification pipeline, i.e., the cross-border transfer intention classifier and the individual transparency element classifiers. To this end, we took advantage of the large-scale compliance assessment method presented in Section 5, which automatically tagged the privacy policies of 10,080 apps using the aforementioned classifiers. A cluster sampling was conducted on these tagged privacy policies to randomly select a subset of 30 privacy policies, while ensured a balanced number of each transparency element. These 30 policies were manually annotated and used as ground truth for the evaluation[9]. Since the actual compliance checking is based on an entire privacy policy, we say that a privacy policy discloses a cross-border transfer practice or a transparency element if at least one policy segment contains them.

Table 6 shows the performance of the classifiers on the 30 unseen privacy policies. As can be observed, the cross-border transfer intention classifier achieves the highest performance in identifying when this practice is disclosed as well as when is not disclosed in a privacy policy. That consistency is particularly important as it minimizes the carry-over of misclassifications into the transparency element classifiers that follow the pipeline. With F-measures ranging from 94.4% to 100% in 3 out of 4 *appropriate measure* classifiers, 90.9% in the *copy means* classifier, and from 85.7% to 100% in the *target country* classifiers, we believe that our approach can be exploited to extract these privacy practices at the level of transparency elements with a high degree of certainty.

---

[9] These annotated privacy policies can be found in the sheet assembled_30_validation.csv available at the replication package.

We examined the misclassifications of the target country (United Arab Emirates) and the explicit consent classifiers, which have the lowest performance. The false negative in the country classifier is because the privacy policy used a non-standard country code (UAE instead of EA), which escapes the proposed approach. The explicit consent classifier did not distinguish between tacit and explicit consent. This is an area that could be improved, perhaps through an effort to extract more positive ground truths and then build a robust ML-based classifier. Nevertheless, the classifier achieves the highest recall, thus avoiding pointing out a wrong compliance issue.

### 4.2 App Behaviour Analysis

This process aims to analyse the behaviour of an Android app and then extract the personal data flows in terms of (i) the type of personal data; (ii) the type of recipient who receives the personal data (*i.e.,* first-party or third-party recipient); and, (iii) the country in which the recipient servers are located. This information feeds the compliance checking process to be compared with the practices extracted from the app privacy policy.

We rely on dynamic analysis to observe the app behaviour and extract its personal data flows. Personal data flows can also be inferred from the app's representations or models by using static analysis [26]–[29]. However, we favour dynamic analysis to prioritise soundness over completeness, as our goal is to extract actual evidence of cross-border transfers carried out by an app. Furthermore, we rely on app network interfaces as sources of behaviour. Previous studies [30] have shown the prevalent usage of network interfaces over SMSs or short-range interfaces such as Bluetooth or NFC by mobile apps to communicate externally. Thus, it is fair to assume that most cross-border transfers occur naturally through the network. Figure 6 sets out the overall data flow extraction process, which is summarized below. Details can be found in [8].

**Configuration.** Based on the Google Play API, we automatically crawl and download the target mobile app (APK) from Google Play Store. Once downloaded, we extract the metadata from the APK digital certificate.

**Stimulation.** Automated stimulation is based on a random strategy provided by the U/I Exerciser Monkey [31], which provides better performance in terms of code coverage compared to other approaches [32].

TABLE 6. PERFORMANCE OF IDENTIFYING THE CROSS-BORDER TRANSFER INTENTION AND THEIR TRANSPARENCY ELEMENTS ON A SET OF UNSEEN PRIVACY POLICIES ($n = 30$).

| Class | Precision | Recall | F-measure | NVP | Specificity | F-measure-negative | +/- Support |
|---|---|---|---|---|---|---|---|
| Cross-border Transfer | 100% | 100% | 100% | 100% | 100% | 100% | 24/6 |
| Transfer & Standard Data Protection Clauses | 93.8% | 100% | 96.8% | 100% | 93.3% | 96.6% | 15/15 |
| Transfer & Adequacy Decision | 89.5% | 100% | 94.4% | 100% | 84.6% | 91.7% | 17/13 |
| Transfer & Binding Corporate Rules | 100% | 100% | 100% | 100% | 100% | 100% | 5/25 |
| Transfer & Explicit Consent | 37.5% | 100% | 54.5% | 100% | 81.5% | 89.8% | 3/27 |
| Transfer & Copy Means | 100% | 83.3% | 90.9% | 96.0% | 100% | 98.0% | 6/24 |
| Transfer & United States | 85.0% | 94.4% | 89.5% | 90.0% | 75.0% | 81.8% | 18/12 |
| Transfer & Canada | 100% | 100% | 100% | 100% | 100% | 100% | 3/27 |
| Transfer & Singapore | 75.0% | 100% | 85.7% | 100% | 96.3% | 98.1% | 3/27 |
| Transfer & Russia | 100% | 100% | 100% | 100% | 100% | 100% | 3/27 |
| Transfer & Japan | 100% | 100% | 100% | 100% | 100% | 100% | 3/27 |
| Transfer & Mexico | 100% | 100% | 100% | 100% | 100% | 100% | 1/29 |
| Transfer & China | 75.0% | 100% | 85.7% | 100% | 96.3% | 98.1% | 3/27 |
| Transfer & Azerbaijan | 100% | 100% | 100% | 100% | 100% | 100% | 1/29 |
| Transfer & Brazil | 100% | 100% | 100% | 100% | 100% | 100% | 3/27 |
| Transfer & Argentina | 100% | 100% | 100% | 100% | 100% | 100% | 1/29 |
| Transfer & Israel | 100% | 100% | 100% | 100% | 100% | 100% | 1/29 |
| Transfer & United Arab Emirates | 0.0% | 0.0% | 0.0% | 96.7% | 100% | 98.3% | 1/29 |
| Transfer & South Korea | 100% | 100% | 100% | 100% | 100% | 100% | 3/27 |
| Transfer & Australia | 75.0% | 100% | 85.7% | 100% | 96.3% | 98.1% | 3/27 |
| Transfer & Belarus | 100% | 100% | 100% | 100% | 100% | 100% | 1/29 |

*The Support column shows the number of ground truths of (+) privacy policies that disclose the corresponding transparency element and (-) the privacy policies that do not disclose it. The binary SVM classifier of cross-border transfer built upon 1-2 grams-based feature vectors after applying stemming and the TF weighting approach.*

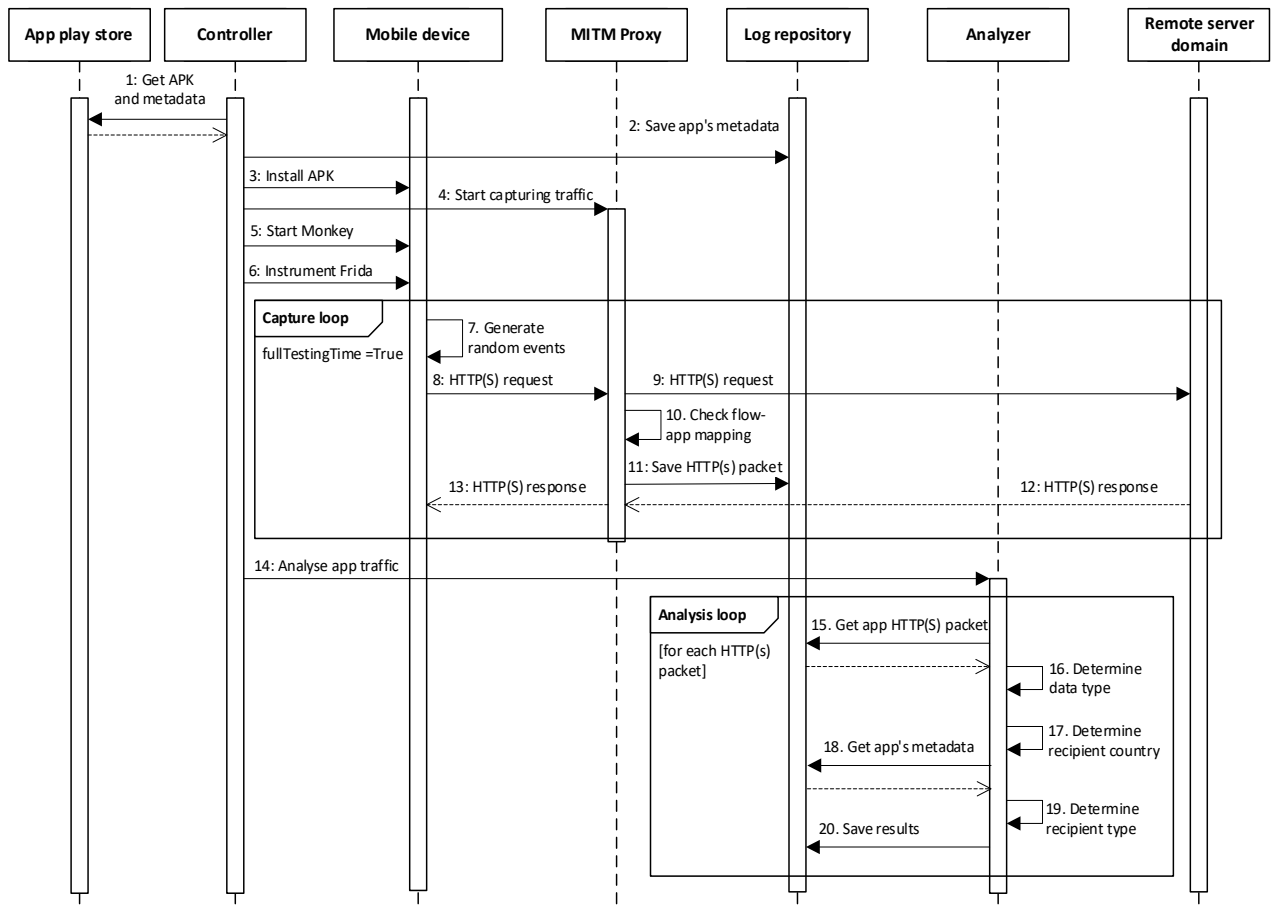

Fig. 6. Overall process to extract cross-border personal data flows from apps.

**Interception.** This component is responsible for capturing the app network traffic and storing it for further analysis. Traffic capture from the device's network interface is built around a man-in-the-middle (MITM) proxy[10], which requires installing a self-signed CA certificate to become trusted. We leveraged Frida[11] to further bypass the most common countermeasures to HTTPS interception e.g. certificate pinning. After configuring the mobile device to connect to the Internet through the MITM proxy, it is possible to capture both HTTP and HTTPS traffic from any app. We further implemented a flow-to-app mapping component to filter in the traffic belonging to the target app since each app is analysed independently.

**Analysis.** This component analyses each app traffic to determine (i) the type of personal data transferred by an app, (ii) the country where the personal data recipient is located, and (iii) the type of recipient who receives the personal data (i.e., first-party or third-party recipient). For (i), we use string searching in the packets payload. It also searches for data encoded in Base64, MD5, SHA1 and SHA256. For (ii), we relied on ipstack API[12] to determine the location of the servers receiving the app's connections. For (iii), we first performed a token matching between a bag of tokens representing the app and a bag of tokens representing the target domain. The former consist of the second-level domain (SLD) and subdomains from the APK name, the organisation name extracted from the digital certificate used to sign the app, and the app name retrieved from the Play Store. The latter consists of the SLD and subdomains of the domain targeted by the traffic. A token matching is then made between the two bags, classifying the domain as a *first-party recipient* if there is at least one token match. Domains not classified as *first-party recipients* were searched in webXray[13]; if found, they were classified as *third-party recipients*. This dataset[14] has been created in the specific context of disclosing personal data to third parties in the web and mobile ecosystem. It maps individual target domains to the owner company and even to parent companies, including the country in which the headquarters are located and service category. Domains not classified as a *first-* or *third-party recipient* were classified as *unknown* and excluded from further analysis.

---

[10] https://mitmproxy.org/
[11] https://frida.re/
[12] https://ipstack.com/
[13] https://webxray.org/
[14] As a further contribution of this study, we added 234 new domains to the dataset maintained by a research community, available at https://github.com/PrivApp/webXray_Domain_Owner_List

## 4.3 Compliance Checking

The final process aims to check whether the apps performing cross-border transfers properly disclose them through their privacy policies. To this end, we consider four consistency types between the app's personal data flows and privacy policy statements, as illustrated below.

### 4.3.1 Full cross-border transfer disclosure

It implies that a privacy policy discloses all *transparency elements* according to the type of cross-border transfer actually carried out by an app. For illustrative purposes, consider the "Viber Messenger" com.viber.voip app, owned by Viber Media. It has been installed +50,000,000 times from the Google Play Store. For marketing purpose, it transfers the AAID (Android Advertisement Identifier) to, *inter alia,* app.adjust.com, whose servers are located in the United States (US). The domain adjust.com is owned by the third-party recipient Adjust, which is based in the US.

The privacy policy of the app includes the statement shown in Figure 7, which fully discloses the cross-border transparency elements. That is, the transfer intention (green); target country (yellow); appropriate safeguard - Standard Data Protection Clause and Binding Corporate Rules (blue); and the means to get a copy of these safeguards (grey). Therefore, in this specific case, we classify this app as a full cross-border transfer disclosure.

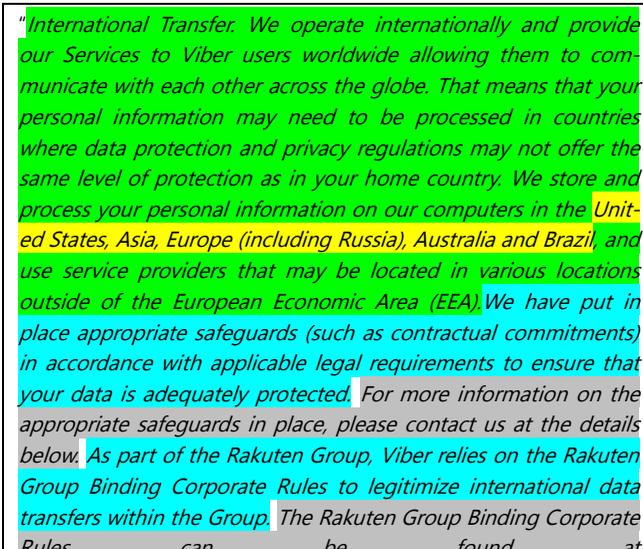

Fig. 7. International transfer statement of the *com.viber.voip* app.

### 4.3.2 Ambiguous cross-border transfer disclosure

It implies that a privacy policy includes only a subset of the *transparency elements* required by the GDPR according to the type of cross-border transfer carried out by the app. The missing *transparency elements* are either included in an ambivalent manner or not included at all. For example, consider the "TapVPN Free VPN" (pm.tap.vpn) app, owned by Smart Media, which has been installed +10,000,000 times from the Google Play Store. For analytics purposes, it transfers the AAID to, *inter alia,* startup.mobile.yandex.net, whose servers are located in Russia. This domain is owned by the third-party recipient Yandex LLC that is based in Russia.

The privacy policy of the app includes the statement shown in Figure 8, which discloses the *intention* to transfer personal data (green). However, it does not reveal the target countries neither *appropriate safeguards* and the *means to get a copy* of such safeguards. Despite this app provider appeals to the consent, the AAID is transferred to startup.mobile.yandex.net before the user interacts with the app for the first time, nullifying any attempt to underpin the transfer by *explicit consent*, as explained in Section 3.

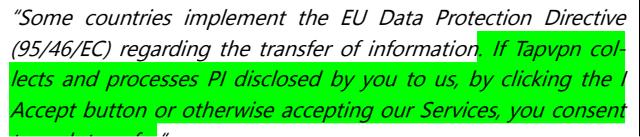

Fig. 8. International transfer statement of the *pm.tap.vpn* app.

### 4.3.3 Inconsistent cross-border transfer disclosure

It implies that a privacy policy includes statements that contradict the cross-border transfers actually carried out by an app. For illustrative purposes, consider the "Kids Learn Professions" (com.forqan.tech.Jobs) app, owned by Jobs Match, that has been installed +10,000,000 times from the Google Play Store. For analytics and advertisement purposes, it transfers the AAID and a Fingerprinting identifier to, *inter alia,* the domain ads.api.vungle.com, whose servers are located in the US. This domain is owned by the third-party recipient Vungle based in the US.

The privacy policy of the app includes the statement shown in Figure 9. It properly discloses all three transparency elements, *i.e.,* the *intention* to transfer personal data (green) to a country (yellow) covered by an *adequacy decision* (blue). However, an international transfer is actually made to the US. Therefore, in this specific case, we classify this app as *inconsistent cross-border transfer disclosure*.

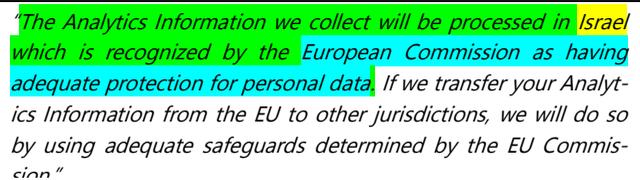

Fig. 9. International transfer statement of the *com.forqan.tech.Jobs* app.

### 4.3.4 Omitted cross-border transfer disclosure

It implies that a privacy policy does not include any *transparency element* when an app performs a cross-border transfer. For example, consider the "PrimalCraft" (com.tellurionmobile.primalcraft) app that has been installed +10,000,000 times from the Google Play Store. It transfers personal data to 13 different third-party recipients, including transfers of the GPS Location and AAID to the domain sdk-android.ad.smaato.net, owned by the advertisement company Smaato and hosted in servers located in the USA. Furthermore, it transfers GPS Location, SSID Wi-Fi information and MAC address information to the domain ad.mail.ru, a social network owned by the company Mail.ru based in Rus-sia. After searching for a policy statement describing at least the intention to make a cross-border transfer, it was not found at all. We, therefore, classify this app as *omitted cross-border transfer disclosure*.

## 5 GOOGLE PLAY STORE APPS ASSESSMENT

To advance the fundamental understanding of GDPR compliance in the Android ecosystem, we carried out a compliance assessment of cross-border transfers of the free, most-popular 10,080 apps from the Google Play Store in Spain. In particular, after defining the experimental environment set-up (Section 5.1), we examined how many apps conducted cross-border transfers and how many of them disclosed properly such practices through privacy policies according to the type of cross-border transfer (Section 5.2), and discuss our findings (Section 5.3).

### 5.1 Experimental Environment

We conducted a controlled experiment by using the assessment method presented in Section 4. Both the Android mobile apps and their privacy policies were downloaded and tested between 20 July and 22 August 2020 from Spain. The apps were installed and tested on five mobile devices: 3 Xiaomi Redmi 7a (API 28) and 2 Xiaomi Redmi 5 (API 25). Each app was run for 10 minutes considering two phases: idle stage (i.e., without user interaction) and active stage (i.e., with user interaction). As illustrated in Figure 10, traffic was captured for 2 minutes without any user interaction (idle stage). Then, a monkey was configured to interact with the application for 8 minutes by generating 5000 events (active stage) in that time span. Before starting the active stage, all permissions requested by the app were automatically granted.

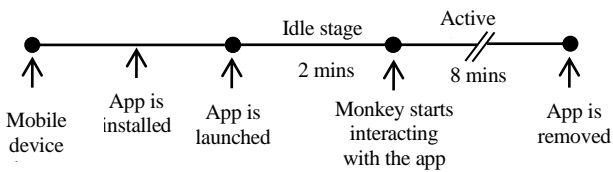

Fig. 10. Testing timeline

On the other hand, we relied on the privacy policy URLs from the console of Google Play Store to download plaintext of privacy policies, as the European Data Protection Board (EDPB) explicitly states that, for apps, the necessary should be made available from an online store prior to download [20]. After retrieving the privacy policy URLs using the Google Play API, we download the content of each policy using the Selenium WebDriver[15] into a headless Chrome browser, allowing the execution of dynamic content such as JavaScript. Non-English privacy policies[16] and their apps were excluded from the analysis.

[15] Selenium Chrome WebDriver available at https://www.selenium.dev/documentation/en/webdriver [Accessed: 23-Mar-2020]
[16] LangDetect was used to determine whether the majority of a privacy policy was written in English. Available at https://pypi.org/project/langdetect/ [Accessed: 23-Mar-2020]

Since the classification models presented in section 4.1 operate based on policy segments, each privacy policy was broken down accordingly, using the full stop as a paragraph separator. Each policy segment was finally pre-processed and then fed into the classification models for extracting the cross-border transfer practices.

**Apps selection.** To have a representative sample of the apps mostly used by EU data subjects, the main criterion guiding the selection of apps was their popularity within an EU country. We relied on Google Play Store's categorisation of the top free apps in Spain to download a set of 10,080 apps[17], which have been highly downloaded as shown in Table 7. These set of apps are distributed across the different categories available on Google Play Store (Fig. 11a). Furthermore, based on the *Issuer Locality* field of digital certificates used to sign the apps, we observe that apps have been supposedly signed by app providers coming from 115 different countries, thus ensuring diversity in the geographical location of app providers. Interestingly, a vast majority of apps (65%) have been signed by non-EU app providers, in particular by providers based in the United States, while only the 8% of apps were signed by EU app providers (Fig. 11b). The remaining 27% of apps do not provide *Locality Issuer* information in digital certificates.

TABLE 7. DISTRIBUTION OF APPS PER DOWNLOADS

| # downloads | # apps (%) |
|---|---|
| < 10.000 | 539 (5.35%) |
| 10.000+ | 997 (9.89%) |
| 50,000+ | 739 (7.33%) |
| 100.000+ | 2,158 (21.41%) |
| 500.000+ | 1,045 (10.37%) |
| 1.000.000+ | 2351 (23.32%) |
| 5.000.000+ | 859 (8.52%) |
| 10.000.000+ | 1,068 (10.60%) |
| 50.000.000+ | 324 (3.21%) |

**App's data flow dataset.** A total of 339,447 data flows generated by the 10,080 apps have been logged. Each log includes the app name, app version, capture stage (idle or active), target domain, target country, and personal data type disclosed (if any). From these flows, 262,253 flows (77.2%) have disclosed at least one of the personal data types to 2,041 unique fully-qualified domain names with 1,309 unique second-level domains (SLDs). These SLDs are hosted on servers located across 38 different countries, 17 EU countries and 21 non-EU countries hosting 165 (12%) and 1,183 (88%) SLDs, respectively. Note that the sum of both exceeds the aforementioned 1,309 unique SLDs because 39 of them are hosted on servers located in both EU and non-EU countries.

**Privacy policy dataset.** Of the 10,080 apps finally selected, 254 (2.5%) have not published their privacy policy URL, while 165 (1.6%), although published, were not reachable at harvest time. The remaining 9,661 (95.9%) privacy policies were fed into the classification models irrespective of whether cross-border transfers were detected during the testing time of their corresponding apps. A total of 355,009 policy segments have been analysed. Of these, 21,122 contain transparency elements of cross-border transfer practices as shown in Table 8. Some transparency elements are redundant within a privacy policy. For example, the cross-border transfer intention is disclosed in 8,408 different policy segments from only 3,808 apps' privacy policies.

[17] Actually, an initial set of 10,470 apps were considered, but 390 were excluded as their privacy policies are not written in English, leaving the 10,080 considered in this study.

TABLE 8. NUMBER OF APPS AND STATEMENTS DECLARING THE TRANSPARENCY ELEMENTS OF CROSS-BORDER TRANSFERS.

| Transparency element | Number of apps | Number of statements |
|---|---|---|
| Data controller/representative | 2,285 (22.6%) | 5,937 |
| Transfer intention | 3,808 (37.7%) | 8,408 |
| Specific target country | 2,808 (27.9%) | 5,159 |
| Existence of EU adequacy decision | 940 (9.32%) | 1,952 |
| Standard Data Protection Clauses | 831 (8.24%) | 978 |
| Binding Corporate Rules | 46 (0.46%) | 54 |
| Explicit consent | 138 (1.37%) | 166 |
| Means to get a copy of safeguards | 370 (3.67%) | 420 |

## 5.2 Results

The number of apps that fulfil the criteria leading to the four types of cross-border transfers is detailed in Figure 12. It was found that *three quarters* (7,579) of the apps transferred some type of personal data during the testing period. From them, a subset of 1,508 apps (15% of total) transferred personal data solely to other EU countries. As explained in Section 3, these do not imply further requirements in terms of GDPR. The remaining 6,071 (60%) apps transferred personal data outside the EU. This subset of apps branches out into three groups that imply different transparency requirements: a small subset of 758 apps transferred personal data to non-EU first-party recipients (Section 5.2.1), a reduced subset of 75 apps transferred personal data to third-party recipients covered by an adequacy decision (Section 5.2.2), and, finally, a substantial subset of 5,665 apps transferred personal data to third-party recipients not covered by an adequacy decision (Section 5.2.3). Note that some apps have performed more than one cross-border transfer type. For example, out of 6,071 apps that transferred personal data to a non-EU country (C1.3.2 in Figure 12), a subset of 365 apps transferred personal data to both non-EU third-party recipients and non-EU first-party recipients. Thus, the sum up of both subsets of apps (5,678 and 758) exceeds its input in 365.

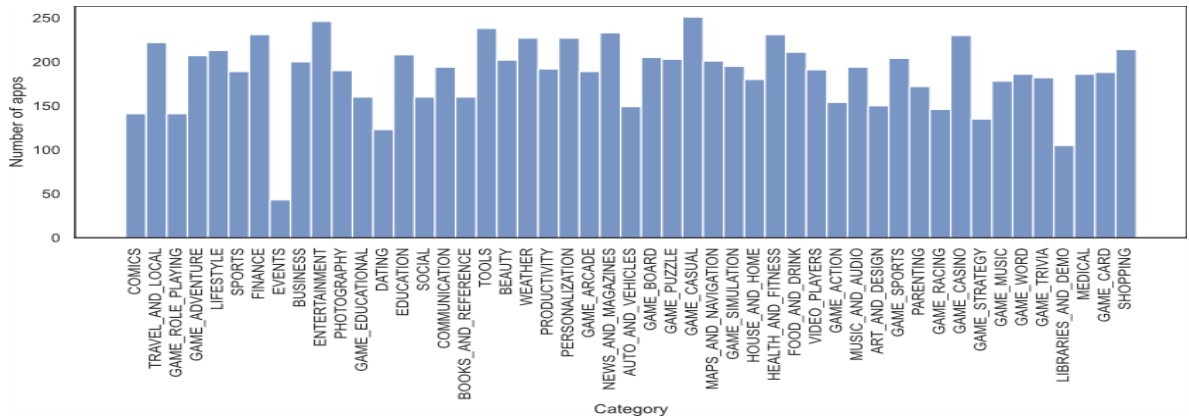

(a)

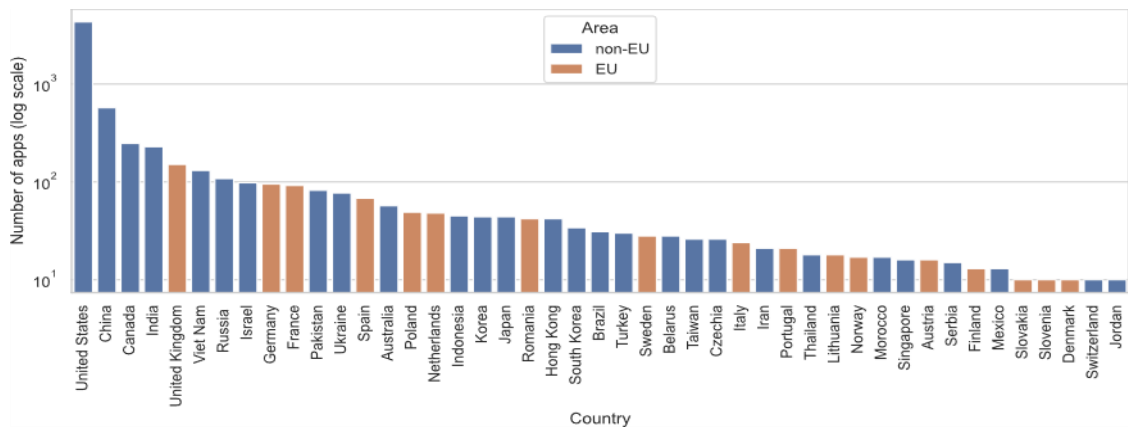

(b)

Fig. 11. Distribution of (a) categories and (b) providers' locality for the apps analysed. Note that (b) is represented in a log-scale axis and due to space limitations only the top-45 countries are shown.

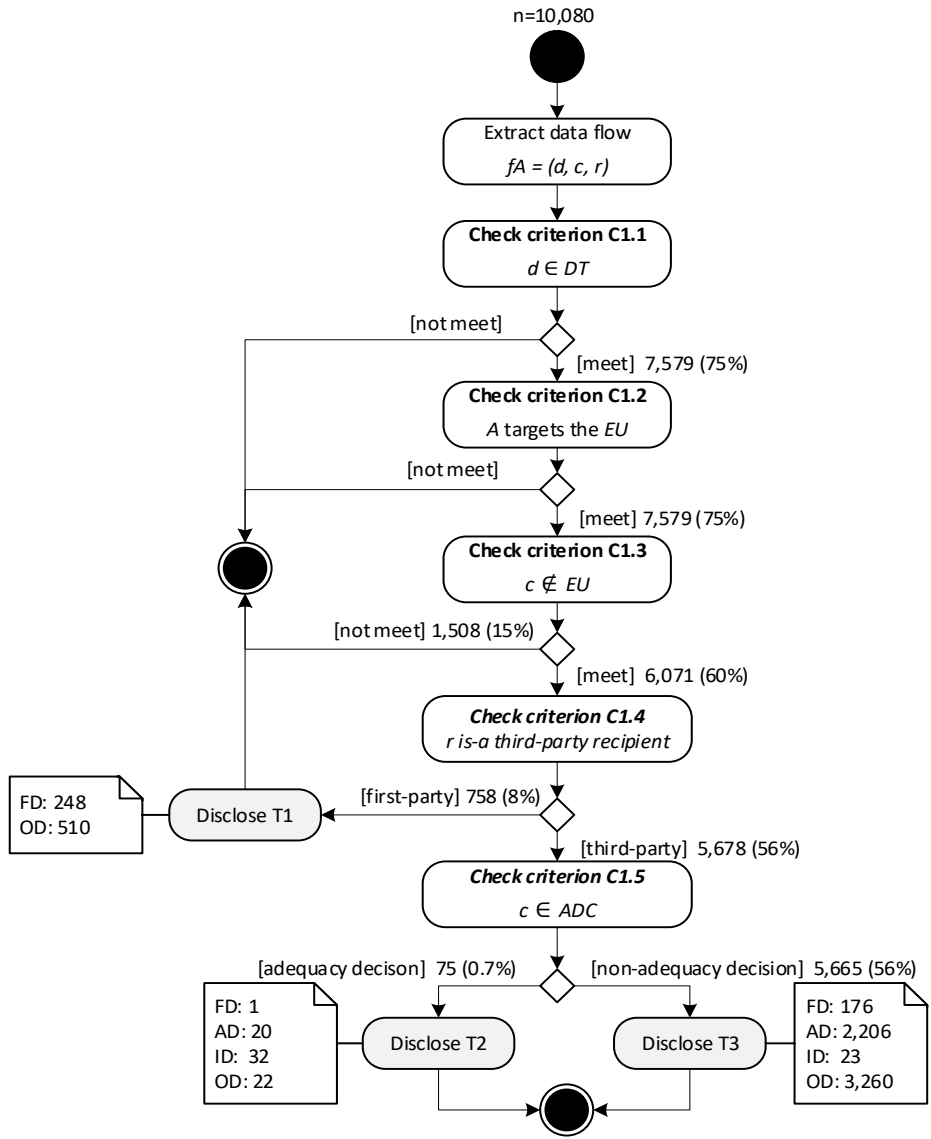

Fig. 12. Number of apps performing cross-border transfers. Each app transfer has been classified as full cross-border transfer disclosures (FD), ambiguous cross-border transfer disclosures (AD), inconsistent cross-border transfer disclosures (ID) or omitted cross-border disclosures (OD). The % is relative to the total number of apps (n=10.080).

### 5.2.1 Transfers to non-EU data controllers

During the testing time, a subset of 758 apps performed cross-border transfers to domains belonging to them but hosted on servers outside the EU[18]. In particular, the United States is the country that hosts the vast majority (89%) of first-party recipient domains. The remaining 11% of these apps host their domains in 13 different countries, including Canada, China, Singapore, Russia, Iran, Japan, India and Australia.

Nearly a third (248) of these apps inform accordingly about an EU representative or data controller[19] and therefore have been classified as **full cross-border transfer disclosures (FD)**, while the remaining 67% **omitted cross-border transfer disclosures (OD)**.

To perform a more in-depth analysis, we relied on the *Issuer Locality* information of each app digital certificate to identify the app provider location and thus distinguishing those that appear to be established within the EU from those established outside the EU. To be sure of the reliability of this assumption, we manually checked the consistency between the supposed issuer country extracted from 40 randomly selected apps and the organisation's country (if any) disclosed in their respective privacy policies, finding only 1 inconsistency[20].

As a result, a remarkable 76% of apps providers (579) would be established outside the EU, compared to a small 7% of apps (129) within the EU, while the *Issuer Locality* of the remaining 17% (129) of apps is missing from their digital certificates. Focusing on the 579 non-EU app providers, a substantial 70% (403) of apps omitted the disclo-sure on an EU representative/data controller in the EU in their privacy policies, thus raising a potential compliance issue. Four key different app providers attitudes can be stemmed after analyzing in-depth some involved privacy policies.

---

[18] Details on non-EU data controller cross-border transfers can be found the *T1_results.csv* available at the replication package.
[19] Strictly speaking, it is only a representative, but we have not noticed that some privacy policies refer to it as an EU data controller.

[20] Details can be found in the *Checking_issuer_locality.csv* sheet available at the replication package.

Several app providers do disclose the intention to perform a transfer outside the EU to their premises in a specific country but appeal to tacit consent as a unique enabling mechanism to perform a cross-border transfer, which is not valid under the GDPR. For example, the *flipboard.app*, a News & Magazine app that has been downloaded more than 50,000,000 times, resorts to the following statement: "*As a California-based company, we store and use personal data outside the EU. By using our websites or submitting your personal data, you consent to such transfer, storing and processing*".

Second, some app providers explicitly state that the underlying service targets a specific country market, other than EU, although disclaims liability for any cross-border transfers of personal data of those who still choose to use them. For example, the *com.hulu.plus*, a popular video stream app that has also been downloaded more than 50,000,000 times, disclaims the following: "*Hulu is headquartered in the U.S. and the Hulu Services are intended for users in the U.S. By viewing any Content or otherwise using the Hulu Services, you consent to the transfer of information to the U.S. to the extent applicable, and the collection, storage, and processing of information under U.S. laws.*".

Some app providers inform about the collection of personal data but omit any applicable regulations at all, so they cannot be expected to designate an EU data representative or controller.

Finally, a minor number of apps (19) still rely on the Privacy Shield framework as an enabling mechanism to perform a cross-border transfer without further constraints, as it was covered by an EU adequacy decision. Yet, as already mentioned, it was invalidated by the EU Court of Justice on July 16, 2020.

### 5.2.2 Transfers covered by adequacy decision

A small set of 75 apps transferred personal data to third-party domains hosted on servers located in countries covered by an EU adequacy decision[21]. In particular, Japan (43 apps), Canada (29), New Zealand (3), and Argentina (1) were targeted by the apps involved. Since all these countries maintain an adequacy decision, which is an assurance mechanism to ensure a protection level equivalent to GDPR, any transfer can be made without any further safeguards. However, as explained in Section 3, three transparency elements should be disclosed to data subjects when such transfers take place, namely transfer intention, names of targeted countries, and the existence of an adequacy decision itself. On balance, we found that 70% of applications disclose to some extent the intention to perform cross-border transfers to countries covered by an adequacy decision, while the remaining 30% omitted it at all.

More specifically, first, only one app discloses all the three transparency elements and therefore have been classified as **full cross-border transfer disclosures (FD).** Second, a subset of 20 apps (27%) has been classified as **ambiguous cross-border transfer disclosures (AD),** as despite they disclose the intention to perform a cross-border transfer, they fail to inform the target countries. The EC has recognized twelve different countries as offering an adequate level of data protection, so it is ambiguous to which countries personal data is transferred. Third, about 43% (32) of apps have been classified as **inconsistent cross-border transfer disclosures (ID)**, because despite disclosing the transparency elements required, there is a disagreement between the countries to which the data are actually transferred and the countries disclosed in their privacy policies. Finally, about 30% (22) of apps omitted disclosure of cross-border transfer practices at all and have been therefore classified as **omitted cross-border transfer disclosures (OD).**

### 5.2.3 Transfers not covered by adequacy decision

The majority of apps (5,665) transferred personal data to third-party recipients located in countries that are not covered by an adequacy decision[22]. In particular, we found that the vast majority of these apps (96%) performed cross-border transfers to the United States. The remaining 4% of apps targeted third-party recipients located in nine different countries: Russia (139), Iran (22), China (20), Singapore (11), South Korea (6), India (4), Australia (1), South Africa (1), and Ukraine (1).

As explained in Section 3, four transparency elements should be disclosed to data subjects when such transfers take place, namely transfer intention, names of targeted countries, appropriate safeguards, and a means to get a copy of such implemented safeguards. However, a reduced 3% (176) of these apps have disclosed the four aforementioned transparency elements through their privacy policies and therefore have been classified as **full cross-border transfer disclosures (FD).**

Besides, less than 1% (23) of these apps do disclose the intention to perform a cross-border transfer but there is a disagreement between the countries actually targeted by transfers and the countries disclosed in the apps' privacy policies. Therefore, they have been classified as **inconsistent cross-border transfer disclosures (ID).**

Also, 36% (2,206) of these apps disclose the intention to perform a cross-border transfer but omit one or more of the other three transparency elements and therefore have been classified as **ambiguous cross-border transfer disclosures (AD).** More specifically, the majority of these apps' privacy policies (1,698) fail to inform on the *appropriate safeguards* or the *means to get a copy* of them. Interestingly, 411 of these apps performed transfers to the United States and still rely on the Privacy Shield framework as an enabling mechanism to perform a cross-border transfer without further constraints. This framework was invalidated by the EU Court of Justice on July 16, 2020, as it does not provide an adequate level of protec-tion, and therefore these apps fall into non-compliant ones. Another remarkable aspect is that several apps (360) disclose the implementation of appropriate safeguards (almost all of them through the establishment of *Standard Data Protection Clauses*) but fail to provide data subjects a means to obtain a copy of these safeguards (*e.g.*, an email or download URL). Also, a concern that arises for all types of cross-border transfers is that the privacy policies of several applications use ambivalent statements, such as "*countries around the world*", *"outside the EEA"* or "*any country in which we do business"*, to refer to the targeted countries.

---

[21] Details on adequacy decision-based cross-border transfers can be found the *T2_results.csv* available at the replication package.

[22] Details on non-adequacy decision-based cross-border transfers can be found the *T3_results.csv* available at the replication package.

Finally, a significant 57% (3,260) of apps **omitted cross-border transfer disclosure** as neither the transfer intention, the recipient countries, nor the appropriate safeguards were disclosed by their privacy policies.

*5.2.4 Who are the third-party recipients?*
We further analyzed the third-party recipients targeted by apps performing cross-border transfers (*i.e,* those 5,678 apps that meet criterion C1.4.2 in Fig. 12). Out of 312 different third-party recipients, as expected, a vast majority (94%) are indeed headquartered across 17 different non-EU countries. Figure 13 shows the top-40 third-party domain owners that have been targeted by apps performing cross-border transfers during our testing. Taking advantage of the classification models presented in Section 4, we examined the privacy policies of the top-110 third-party recipients[23].

Around 12% (13) of these third-party recipients, which collectively have been targeted by 1,551 apps, do not even report the intention to perform a cross-border transfer.

A significant 52% (57) of them, which overall have been targeted by 6,852 apps (note that certain apps may transfer data to more than one third-party recipient), are still appealing to the Privacy Shield framework[24] as an enabling mechanism to perform a cross-border which is not valid at all. More importantly, 30 of them use this framework as a unique assurance mechanism becoming potentially non-compliant.

Unlike mobile apps, which ambiguously disclose the target countries, we positively observe that a significant 78% (86) of third-party recipients do disclose this information explicitly. However, only half of them (40) inform on the appropriate safeguards implementations that enable a cross-border transfer.

We also emphasize that the aforementioned results correspond to the available privacy policies. However, in some cases, finding the privacy policies of third-party recipients can be a challenging process, as they are not centralised like mobile apps. Several of them are not available at all, which denotes the poor transparency practices of the services that are being embedded by app developers. Just to give an example, the *globalcampaign-tracker.com* and *aawrnstrk.com* domains only show a short banner of 144 words. Information about privacy practices, including cross-border transfer practices, is practically non-existent, but their underlying services have been bundled in some way into 26 different apps that transferred personal data to servers outside the EU.

### 5.3 Discussion
**A substantial 56%[25] of analysed apps are potentially non-compliant with the GDPR cross-border transfer requirements**. Despite efforts to ensure cross-border transfers through GDPR, the results clearly reveal that there is still a very significant gap between what app providers and third-party recipients do in practice and what is intended by GDPR. The results show that 56% of popular apps in an EU country are potentially non-compliant with GDPR cross-border transfer requirements. In particular, 32% of mobile apps do not disclose these practices *at all*, while the remaining 26% partially disclose them in an ambiguous and/or inconsistent way.

We argue that two main concerns inherent to mobile apps, consistent with previous work [1], impact on compliance issues for cross-border transfer and leave a long way to go, namely app providers unawareness and lack of transparency from third-party services.

First, app providers may be unaware of the cross-border transfers performed by third-party libraries they embed. The apps studied transferred personal data to third-party recipients, who provide a variety of services predominantly related to advertising and analytics. On average, each app performed international transfers to 2.35 different third-party services. These services often require third-party libraries (TPL) to be embedded in the app code. While app developers may be expected to completely understand the practices of these TPLs before including them in their apps, the evidence indicates that this is not always the case [1].

Second, some third-party services lack transparency on cross-border transfers. TPLs are expected to transparently inform their cross-border transfer practices through privacy policies or terms of service. Thus, app providers who embed them could, in turn, include such practices in their privacy policies. Some TPLs actually require developers to explicitly disclose their integration into the app's privacy policy and to obtain explicit consent as per GDPR [33]. However, as revealed, other TPLs are either ambiguous or do not even provide a privacy policy. Specifically, 12% of the top-110 third-party services do not provide information on their international transfer intentions, and the 36% that do provide it fail to do it transparently.

**Explicit consent is a fair enabler of cross-border transfers, but it is being misused.** In the absence of an *adequacy decision* or any *appropriate safeguards*, the *explicit consent*[26] could also enable cross-border transfers. Nevertheless, *explicit consent* requires a clear affirmative action of the data subjects, *e.g.*, ticking a box, to be obtained after providing **precise details** of the international transfers. As such, explicit consent removes the possibility of using the dark pattern of pre-ticked boxes or tacit consent.

---

[23] Details on the results of the top-110 third-party recipient privacy policies can be found in the *TPL-110.csv* sheet available at the replication package, and the full dataset of third-party recipients at https://github.com/PrivApp/webXray_Domain_Owner_List
[24] These privacy policies were downloaded in December 2020.

[25] Note that that some apps have performed more than one type of cross-border transfer. Therefore, an app has been classified as fully compliant only if all individual transfers have been classified as full cross-border transfer disclosures. Detailed results of each app can be found at the replication package.
[26] GDPR Art. 49

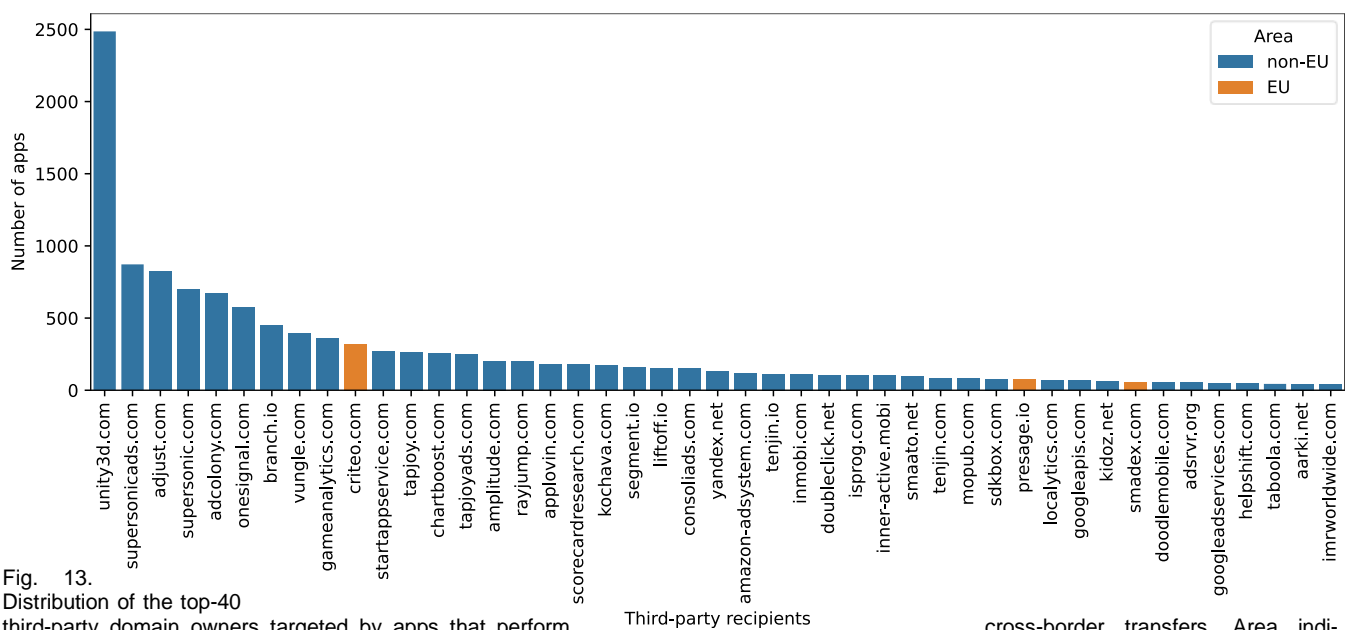

Fig. 13. Distribution of the top-40 third-party domain owners targeted by apps that perform cross-border transfers. Area indicates the headquarters country.

To observe the prevalence in disclosing (explicit) consent as an enabler of cross-border transfers, we further analyzed the apps that meet criterion C1.4.2 (Fig. 12)[27]. In particular, we observed 1,826 privacy policies that disclose the intention to perform a cross-border transfer but not the appropriate safeguard implementation. The result is that around 30% (555) of them resorts to the consent as an enabler of cross-border transfers. We randomly selected a subset of 50 privacy policies and examined the consent-labelled statement. Interestingly, in all cases, app providers appealed to *tacit consent* stating that the usage of the app by the data subject implies a consent to transfer personal data outside the EU. In this line, we observed that most of them (458) performed cross-border transfers during the idle testing stage, *i.e.,* before the user interacts with the app, thus nullifying any attempt to underpin the transfer by *explicit consent*. As already mentioned, while the consent is, in fact, a GDPR legal basis to perform a cross-border transfer, it requires a clear affirmative action of data subjects consenting it before it takes place [3].

**Automated means for compliance assessment are key.** The need for automated methods and tools to evaluate privacy requirements are essential in an evolving regulatory landscape. During the course of this study, we coincidentally experienced a significant change in GDPR cross-border transfer requirements. In 2016, the EU-US Privacy Shield was introduced as a limited adequacy decision to allow the transfer of personal data to US-based third party recipients that were certified under the terms of this framework. US-based app providers extensively used this framework, as evidenced in this study, which allowed these providers to legally target EU data subjects and transfer personal data between the EU and the US. However, this framework was invalidated by the Court of Justice of the EU on July 16, 2020 [34]. The automated approach presented in this paper has enabled a survey on compliance with GDPR cross-border transfers, finding the still prevalent use of the EU-US Privacy Shield in the privacy policies of both apps (Section 5.2.3) and third-party recipients (Section 5.2.4) despite its invalidity. This shows the relevance of our approach in a landscape that is constantly evolving.

In the same line, fully automated methods and techniques offer a viable alternative for large-scale assessment of apps compliance with different requirements. The open model of the Android mobile ecosystem allows a vast number of apps to be offered globally by developers everywhere. Hence, manual review by control authorities or distribution platform operators is impossible. The mean time for downloading and tagging cross-border transfer practices in a privacy policy was 10 seconds, so with a single instance up to 8,640 privacy policies can be tagged in 24h. On the other hand, although in this study we have combined the ML-based classification models with dynamic analysis techniques to obtain evidence of the actual execution of apps, these can be combined with static analysis techniques [10] for greater coverage, or with both static and dynamic, as needed. As such, we strongly believe that our approach can be leveraged for large-scale compliance scrutiny of cross-border transfers.

**A supporting tool for developers.** Whilst our proposed method mainly aims to support audits by authorities or distribution platforms, such as Google Play Store, it can also support app developers in ensuring the compliance of their apps or the third-party libraries they embed in their apps. It should be recognized that the mobile application ecosystem is a mix of formal requirements established by prevailing regulations such as GDPR, along with informal developer training. App developers can range from hobbyist to experienced professionals in large companies. As found in previous work [7], while large companies may be able to form multidisciplinary teams to enforce legal requirements, small business developers may struggle to understand the privacy and data protection implications of their code. We consider that the necessary multidisciplinary knowledge, including evaluation criteria supported by legal and not only technical interpretations, can be simplified into indicators that can be checked automatically. This paper, as well as other related work [9], [10], proved that (at least part of) such multidisciplinary knowledge can be embedded into GDPR automated assessment pipelines. While these automatic approaches do not act as *infallible judges*, they have the potential to alert developers about possible non-compliance issues.

---

[27] Details on the declaration of consent performed by these privacy policies can be found in the *T3_results.csv* sheet available at the replication package.

## 6 THREATS TO VALIDITY

**Construct validity.** The classification models build upon the Corpus IT-100, which is an annotated dataset of legal requirements laid down in the GDPR. Therefore, there is a risk that such a Corpus do not reflect the construct under study when moving legal requirements to the technical domain. To mitigate this threat, the elaboration of the Corpus IT-100 was undertaken by privacy and data protection experts who comprehensively guided the building of the annotation process, annotation scheme, and the simplified assumptions used during the annotation process of cross-border transfer practices in privacy policies.

**Internal validity.** If policies in languages other than English were excluded, a bias towards the evaluation of non-EU based applications could occur. Surprisingly, we observed that the providers of the applications mainly include privacy policies in English. Only 390 apps (3.7%) published non-English policies exclusively, and only 80 were in Spanish (0.76%).

Moreover, our automated privacy policy analysis approach, like any approach based on statistical learning, exhibits problems of misclassification that should be taken into account. Thus, as pointed out in the different results in section 4, although high F-measure values are achieved (from 85.7% to 100%) there is the possibility of a small number of other classifications. All in all, while it certainly does not act as *infallible judges*, we highlight the performance of the cross-border transfer intention classifier which does not exhibit any misclassification in a subset of randomly selected privacy policies, demonstrating its potential to alert stakeholders of potential non-compliance issues.

**External validity.** Since the current implementation of the assessment method is built upon dynamic analysis techniques, it inherits the same limitation faced by them. The use of non-standard encodings mechanisms [35], unusual TLS certificate pinning implementations [36], and sub-optimal coverage of app execution paths [37] are some particular open orthogonal challenges to our proposal, which can generate false negatives. Therefore, it cannot ensure completeness and the results of fully compliant apps should not be misleading generalized. The fact that we have not observed a cross-border transfer during our testing period does not mean that an app will not definitely do so if its developers, *e.g.*, use customized encoding mechanisms.

All in all, potential false negatives do not put at risk the validity of the results of non-compliant apps, which is in fact remarkably high (56%). The strength of dynamic analysis techniques is that evidence of non-compliant apps stem from real app behaviour and do not generate false positives. Therefore, we consider that our proposal, as well as the results, are valuable for app providers, app distribution platforms such as Google Play Store, and supervisory authorities to detect the lower bound of non-compliance issues with GDPR cross-border transfers.

## 7 CONCLUSION AND FUTURE WORK

In this work, we presented a fully automated method to assess compliance of mobile apps with the cross-border transfer requirements established by the GDPR. With an F-measure ranging from 85.7% to 100% in identifying the different cross-border transfer transparency elements, our approach can be exploited to extract these privacy practices with a high degree of certainty and at scale.

Also, we applied the automated compliance assessment method to determine the extent to which the apps from the Google Play Store comply with the cross-border transfer requirements of the GDPR. After evaluating the top-free 10,080 apps from the Google Play Store in Spain, the results revealed that there is still a great gap between what app providers and third-party services do in practice and what is intended by GDPR. Notably, *5,646 (56%) apps* failed (completely or partially) to comply with the regulations, either because their privacy policies include ambiguous or inconsistent disclosures about cross-border transfers, or they simply omit them. In addition, the results of analysing the privacy policies of the top-110 third-party services, which were collectively targeted by 6,852 apps, revealed a significant 52% of them still relying on the Privacy Shield Framework as an enabling mechanism to perform cross-border transfers, almost six months after it was invalidated.

In a complex and evolving regulatory landscape automated methods and tools to evaluate privacy requirements are essential to several stakeholders, including supervisory authorities, distribution platforms and developers. Our current efforts are aimed at extending the analysis of privacy policies disclosed in other languages. Likewise, we aim to extend our method to address other requirements of the GDPR that have not yet been subject of research efforts, in particular those related to *automated decision-making*.

## ACKNOWLEDGMENT

This research has been partially supported by the CLIIP project (grant reference APOYO-JOVENES-QINIM8-72-PKGQ0J) funded by the Comunidad de Madrid and Universidad Politécnica de Madrid under the V-PRICIT research programme 'Apoyo a la realización de Proyectos de I+D para jóvenes investigadores UPM-CAM', and by the Escuela Politécnica Nacional in Ecuador.

Jose M. del Alamo (jm.delalamo@upm.es) is the corresponding author.